\documentclass{PoS}
\usepackage{graphicx}
\usepackage{latexsym,amsmath,amssymb,lmodern,float,url}
\usepackage{color}
\usepackage{microtype}
\usepackage{slashed}
\usepackage{multirow}

\newcommand{\be}{\begin{equation}}
\newcommand{\ee}{\end{equation}}
\newcommand{\bea}{\begin{eqnarray}}
\newcommand{\eea}{\end{eqnarray}}
\def\beqs#1\eeqs{\be\begin{split} #1 \end{split}\ee}

\newcommand{\figu}[1]{Fig.~\ref{fig:#1}}

\newcommand{\ba}{\begin{array}}
\newcommand{\ea}{\end{array}}

\newcommand{\im}{\mathop{\rm{Im}}}
\newcommand{\as}{\langle\sigma\rangle}

\def\av#1{ \left\langle #1 \right\rangle }

\def\Im {\mathop{\hbox{Im}}}

\title{Flowing Gauge Theories: Finite-Density $QED_{1+1}$}

\ShortTitle{Flowing Gauge Theories: Finite-Density $QED_{1+1}$}

\author{\speaker{Henry Lamm}\\
        University of Maryland\\
        E-mail: \email{hlamm@umd.edu}}

\abstract{Finite-density calculations in lattice field theory are typically plagued by sign problems. A promising way to ameliorate this issue is the \emph{holomorphic flow} equations that deform the manifold of integration for the path integral to manifolds in the complex space where the sign fluctuations are less dramatic. We discuss some novel features of applying the flow equations to gauge theories and present results for finite-density $QED_{1+1}$.}

\FullConference{The 36th Annual International Symposium on Lattice Field Theory - LATTICE2018\\
		22-28 July, 2018\\
		Michigan State University, East Lansing, Michigan, USA.}

\begin{document}

\maketitle

Observables in non-perturbative field theory are typically calculated via stochastic methods and importance sampling.  Alas, at finite density the action is complex, resulting in a non-real Boltzmann factor $e^{-S}$ that rapidly oscillates as a function of field variables.  This presents a challenge due to strong cancellations between regions. This {\em sign problem} is the main roadblock to the lattice QCD study of dense matter.

If instead of evaluating the path integral over real fields, but as an integral over a chosen integration contour in complex field space, the sign problem can be ameliorated. Cauchy's theorem guarantees that, for classes of contours, the value of the path integral is unchanged.
The original suggestion~\cite{Cristoforetti:2012su,Cristoforetti:2013wha,Cristoforetti:2013qaa,Scorzato:2015qts}  was to use a certain combination, $\mathcal{M}_{\mathcal{T}}$, of {\em Lefschetz thimbles}.  To avoid the difficulties of this method, in~\cite{Alexandru:2015sua}, a so-called {\em generalized thimble method} (GTM) was proposed, in which a manifold $\mathcal{M}_T$ is obtained by continuously deforming $\mathbb R^N$ via the \emph{holomorphic flow equation}:
$\frac{\mathrm d \phi_i}{\mathrm dt} = \overline{\frac{\partial S}{\partial \phi_i}}$.
Evolving every point of $\mathbb{R}^N$ for a ``time" $T$, $\mathcal{M}_T$ is obtained that i) yields equivalent results to the original space and ii) approaches $\mathcal{M}_{\mathcal{T}}$ in the limit $T\rightarrow\infty$, improving the sign problem.  Other manifolds have also been used to reduce computational costs~\cite{Alexandru:2017czx,Mori:2017pne,Alexandru:2018fqp,Mori:2017nwj,Bursa:2018ykf,Alexandru:2018ddf}.

 Apply these methods to gauge theories has developed slowly with only $0+1$ dimensional models, one-plaquette models~\cite{Schmidt:2017gvu} and $QED_{1+1}$~\cite{Alexandru:2018ngw} being studied. Gauge invariance brings a host of conceptual issues, which must be understood in the context of GTM. Unlike previously-studied models, the thimble decomposition in $QED_{1+1}$ is not well-defined; despite this GTM yields correct results and improves the sign problem.
  
The thermal expectation value of an observable $\mathcal O$ are given by\be\label{eq:expectation-M}
\left<\mathcal O\right> ={\int_{{\mathbb R}^N}\mathcal D\phi\;\mathcal O e^{-S} \over \int_{{\mathbb R}^N}\mathcal D\phi\; e^{-S}} = \frac{\int_{\mathcal M} \mathcal D\phi\;\mathcal O[\phi] e^{-S[\phi]}}{\int_{\mathcal M} \mathcal D\phi\;e^{-S[\phi]}}\,.
\ee
where $\mathcal M$ is a manifold in complex field space.  The average sign is
\be\label{eq:averagesign}
\as\equiv \left<e^{-i S_I}\right>_{S_R} = \frac{\int \mathcal D \phi\; e^{-S}}{\int \mathcal D \phi\;e^{-S_R}}\,.
\ee
Inspecting this expression leads to a conceptual understanding of how the sign problem is reduced.  The numerator has a holomorphic integrand, thus independent of $\mathcal{M}$. However, the denominator's integrand is \emph{not} holomorphic and $\as$ thus depends on $\mathcal{M}$.  We parametrize every point $\tilde\phi_T(\phi)$ on $\mathcal M_T$ by the real coordinates in $\phi$.
Using $\phi$, the expectation values can be written as
\begin{align}\label{eq:Oflow}
\left<\mathcal O\right>=
\frac{\int_{\mathcal M_T} \mathcal D\tilde\phi\;\mathcal O[\tilde\phi] e^{-S[\tilde\phi]}}{\int_{\mathcal M_T} \mathcal D\tilde\phi\;e^{-S[\tilde\phi]}}=
\frac{\int_{\mathbb R^N} \mathcal D\phi\;\mathcal O[\tilde\phi_T(\phi)] e^{-S[\tilde\phi_T(\phi)]} \det J}{\int_{\mathbb R^N} \mathcal D\phi\;e^{-S[\tilde\phi_T(\phi)]}\det J}
\end{align}
where we introduce the Jacobian matrix given by $J_{ij} =\frac{\partial \left(\tilde\phi_T\right)_i}{\partial\phi_j}$.

In the limit $T\rightarrow\infty$, most fields acquire large $S_R$ and decouple from the path integral. In the real-field parameterization, the path integral's support comes from the points that flow into small regions near the critical points. These regions are stretched by flow and generate $N$-real-dimensional surfaces around the critical points. In other words, in the large $T$ limit, $\mathcal M_T$ is the union of approximate thimbles attached to these critical points. Since $S_I$ is constant with flow, so the variation of $S_I$ on an approximate thimble is small. Consequently the sampled fields has a small phase variation, hence a milder sign problem\footnote{Provided the residual phase from $\im\det J$ and cancellations \emph{between} thimbles are negligible.}. 

As a demonstration of the generalized thimble method in a gauge theory, we study a three-flavor $QED$ in 1+1 dimensions.  The action is
\begin{equation}
	S = \frac 1 {g^2} \sum_r \left(1 - \cos P_r\right) - \sum_a \ln \det D^{(a)}
\end{equation}
where $D^{(a)}$ denotes the fermionic matrix for flavor $a$, and $P_r$ denotes the primitive plaquette with $r$ at the lower-right corner $P_r \equiv A_1(r) + A_0(r+\hat x) - A_1(r+\hat t) - A_0(r)$.  Above $\hat{t}$ and $\hat{x}$ are the unit vectors in time and space direction.
We discretize the fermion action using the staggered formulation.
The Kogut-Susskind staggered fermion matrix for flavor $a$ is given by
\begin{equation}
	D^{(a)}_{xy} = m_a \delta_{xy} + \frac 1 2 \sum_{\nu\in\{0,1\}}
	\big[
		\eta_\nu e^{i Q_a A_\nu(x) + \mu \delta_{\nu 0}}\delta_{x+\hat\nu,y}- \eta_\nu e^{-i Q_a A_\nu(x) - \mu \delta_{\nu 0}} \delta_{x,y+\hat\nu}
	\big]
	\text.
\end{equation}
All flavors have the same mass $m_a = m$ and chemical potential $\mu_Q$ but with charge assignments $Q_1=Q_+=+2$, and $Q_{2,3}=Q_-=-1$.

 Previous arguments that GTM improves the sign problem relied upon $\mathcal{M}_T$ approaching the Lefschetz thimbles at $T\rightarrow\infty$.  Where no such unique thimble decomposition exists, it is not obvious what the large-$T$ limit of $\mathcal M_T$ is, and whether the sign problem is reduced. 

In $QED_{1+1}$ dimensions, there are two separate obstacles to defining a unique thimble decomposition: critical points are degenerate (the action does not change along gauge orbits), and lines of flow may connect one critical point to another (Stokes' phenomenon). 

A Lefschetz thimble is defined from an isolated critical point $z_c$ as the union of all solutions $\lim_{T\rightarrow -\infty} z(T) = z_c$. For a holomorphic function of $N$ complex variables, each isolated critical point is a saddle point with $N$ stable and $N$ unstable directions; therefore, each Lefschetz thimble is an $N$-real-dimensional surface. However, in a gauge theory, all critical points form gauge orbits and therefore unisolated.

For an abelian theory like $QED_{1+1}$, the gauge degeneracies can be resolved by gauge-fixing.  In the complexified field space a general gauge transformation is given by 
\begin{equation}
	A_{\mu}(x) \rightarrow A_{\mu}(x) + \alpha(x+\hat\mu) - \alpha(x)
\end{equation}
where $\alpha(x)$ is any complex-valued function on the lattice points and $\hat \mu$ denotes the unit vector along the direction $\mu \in \{0,1\}$. With $V$ lattice sites, the complexified field space is $\mathbb C^{2V}$. The gauge orbit of any configuration $A_{x,\mu}$ is 
obtained by adding to it all vectors of the form $\alpha(x+\mu)-\alpha(x)$, for every $\alpha(x)$ implying a $(V-1)$-dimensional space since $\alpha(x)=\alpha$ is trivial.  Every $A_\mu(x)$ can be decomposed $A_\mu(x)=A_\mu^\perp(x)+A_\mu^\parallel(x)$, with $A_\mu^\parallel(x)$ parallel to the gauge orbit and $A_\mu^\perp(x)$ 
orthogonal to it.
We can choose $A_\mu^\perp(x)$ as the representative of $A_\mu(x)$ in every gauge orbit. In this way, we decompose the original real configuration space as $\mathbb R^{2V} = \mathcal M_0 \oplus \mathcal G$, where $\mathcal M_0 = \mathbb R^{V+1}$ is the gauge-fixed space of $A_\mu^\perp(x)$, and $\mathcal G$ is a single gauge orbit.
The gradient $\overline{\frac{\partial S}{\partial A}}$ is orthogonal to the gauge orbits so
a gauge-fixed slice, defined by constant $A_\mu^\parallel(x)$, is invariant under the flow.  Critical points are isolated in each slice, and a unique definition of the thimble decomposition exists. The behavior of flow can be understood by considering each slice.  Since gauge fixing and flow commute, flowing the entire gauge-free integration domain is $\mathcal M_T\oplus \mathcal G$. In simulations, no gauge fixing is performed, so they perform a random walk in $\mathcal G$.

\begin{figure*}[t]
	\includegraphics[width=\linewidth]{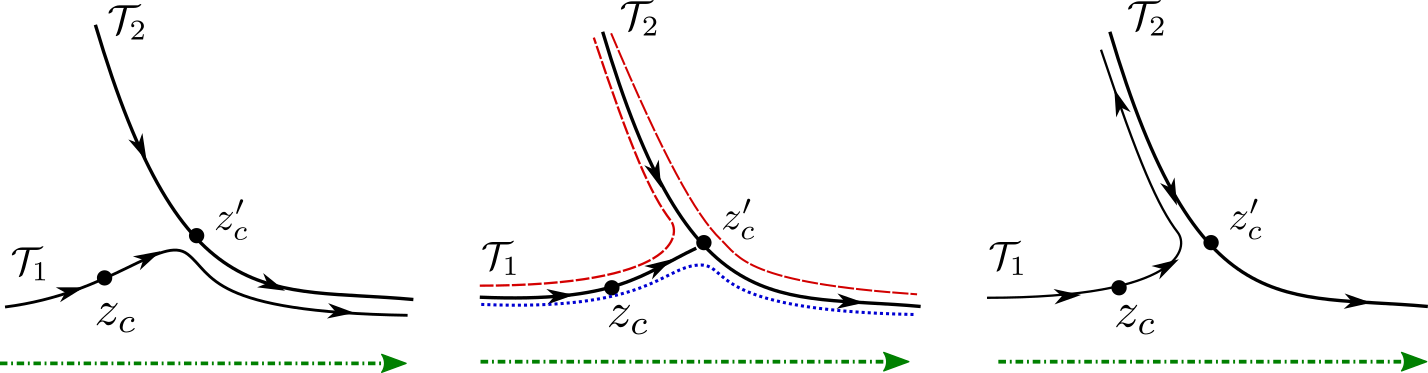}
	\caption{A schematic view of Stokes' phenomenon. The original integration contour is given by the dash-dotted green line. The solid dots denote the critical points, solid lines denote different thimbles, and the arrows denote the orientation of the thimbles. As the argument of $S$ changes (from left to right), the thimble decomposition jumps from $\mathcal T_1$ (left) to $\mathcal T_1$+$\mathcal T_2$ (right). When Stokes' phenomenon occurs (center) there is no unique thimble decomposition, which we indicate by two paths of integration given by the red dashed and blue dotted lines.
\label{fig:stokes}}
\end{figure*}

The second difficulty encountered in the decomposition is Stokes' phenomenon, where a flow line connects two critical points.  The left and right panels of \figu{stokes} show the flow assuming adding a small $i \epsilon$ to the action with opposite signs. In each case, the thimble decomposition is well-defined, but $\mathcal{M}_{\mathcal{T}}$ equivalent to the real domain is different.  Stokes' phenomenon affects the behavior of the holomorphic flow. In the case of $QED_{1+1}$, it occurs at all $\mu$ and $g$.  While Stokes' phenomenon does not cause any discontinuity in the flowed manifolds, it does produce undesirable ``bumps''. As $T$ increases, a bump is created near the origin.  The effect can be understood by considering the $T\rightarrow\infty$ limit, where it is maximized. In this limit, the bump travels up one half of the imaginary axis, and directly back down again, producing a closed contour of integration. The sum of these two segments cancel but the average sign is decreased by their presence.

To separate the effect of the phase fluctuations on $\mathcal M_T$ and the fluctuations induced by approximating $\ln\det J$ with $W$,
we write the reweighting factor $\Delta S = i\Im S_\text{eff} + \Delta J$.  
The phase factor $\exp(-i\Im S_\text{eff})$ is a pure phase.
The real factor $\Delta J$, with $\exp(-\Delta J)=|\det J|/|W|$, is necessary to correct for using $W$ instead of $\det J$ in the Monte-Carlo process.

The speed-up from GTM  compared to real-plane calculations is given in terms of: $t_\text{config}$, the wall-clock time required to generate a configuration; $\langle\sigma\rangle$, the average sign; and $\Sigma$, the {\em statistical power} defined via $w=e^{-\Delta J}/\langle e^{-\Delta J}\rangle_{S_{\rm eff}'}$.
The expression for $\Sigma$ is
\begin{equation}
 \Sigma\equiv\frac{\langle w\rangle_{S_{\rm eff}'}}{\langle w^2\rangle_{S_{\rm eff}'}}=\frac1{\av{w^2}_{S_\text{eff}'}}.
\end{equation}
$\Sigma$ is bounded between $1/n_\text{config}$ and $1$ and is an estimate of the fraction of configurations that effectively contribute. A small value of $\Sigma$ indicates that the averages are dominated by a small number of configurations and, consequently, more configurations need to be used for a reliable estimate.
  
\begin{figure*}[t]
 \includegraphics[width=0.48\linewidth]{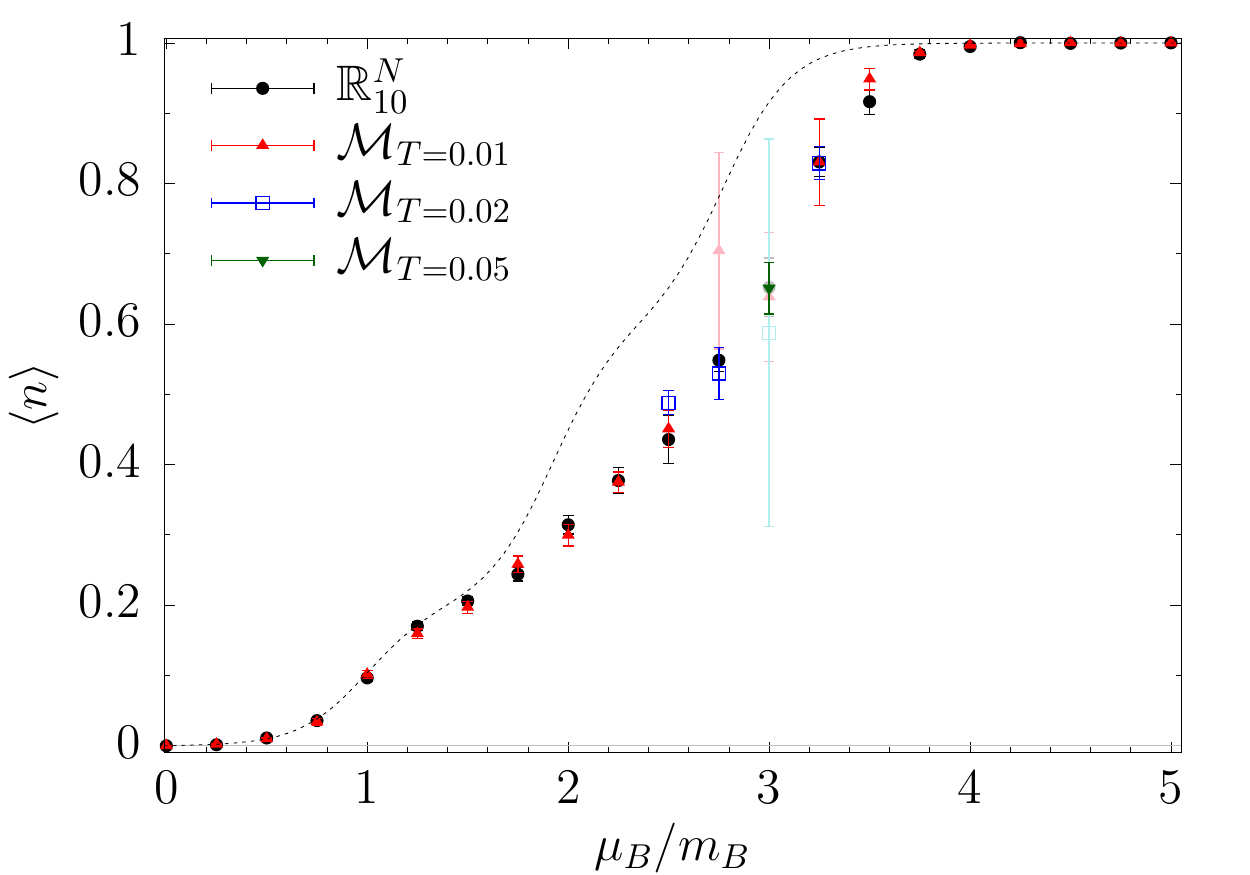}
 \includegraphics[width=0.48\linewidth]{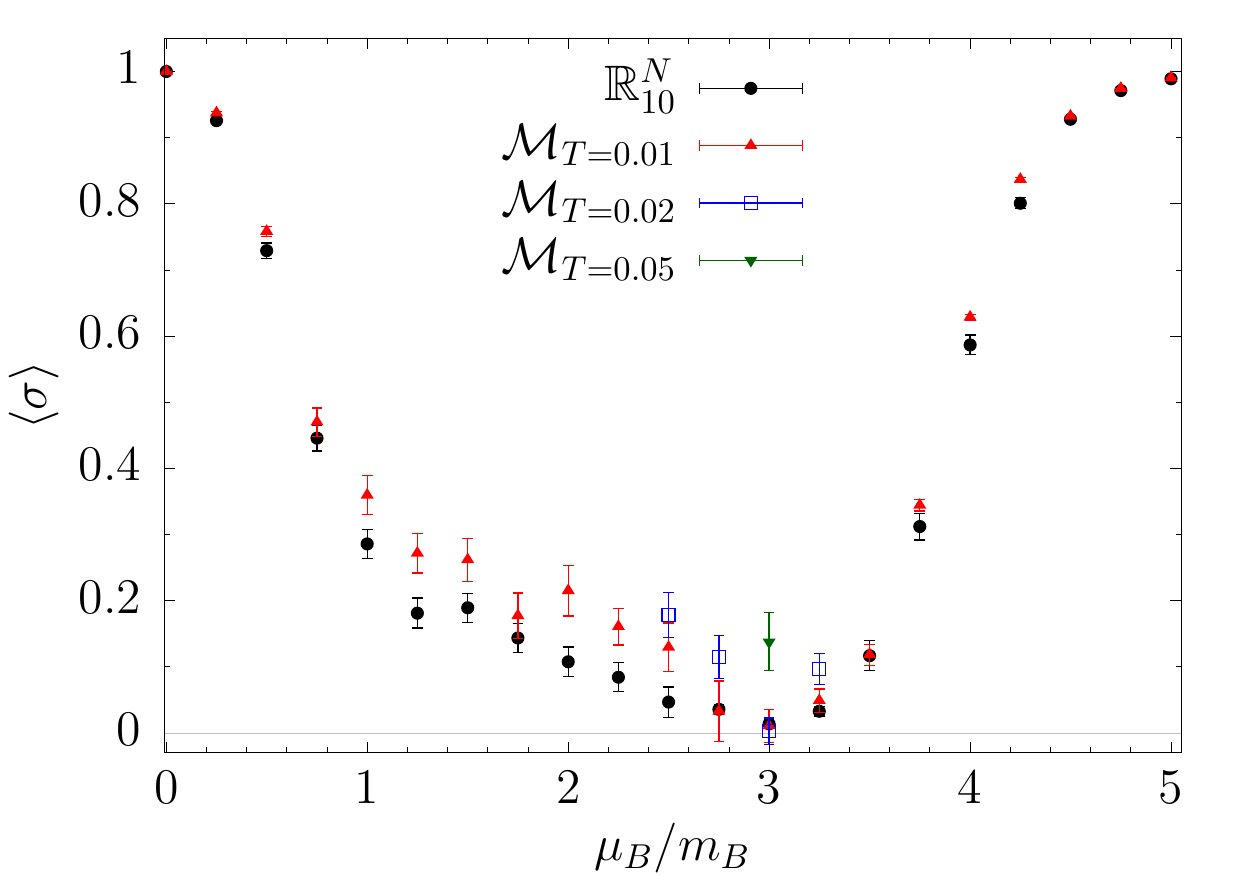}
\caption{\label{fig:1010}Density $\langle n\rangle$ and average sign $\av{\sigma}$ as a function of $\mu_B/m_B$ for staggered fermions on a lattice of size $10\times10$.  The dashed curve represents the free baryon gas with the same mass.}
\end{figure*}
  
A figure of merit $h_T$ for a fixed flow time $T$ can be defined as $h_T= \frac{\langle\sigma\rangle^2\Sigma}{t_\text{config}}$.  A ratio $h_{T_1} / h_{T_2}$ estimates the relative speed-up of flow time $T_1$ over flow time $T_2$.  $\langle\sigma\rangle$ should increase with $T$ because flowing improves the relative sign, but $t_{\rm config}$ also increases because longer flow requires more computational time.  Due to the use of $W$, an approximate Jacobian that is computationally faster, the statistical reweighting plays a nontrivial part. 

\begin{figure*}
 \includegraphics[width=0.48\linewidth]{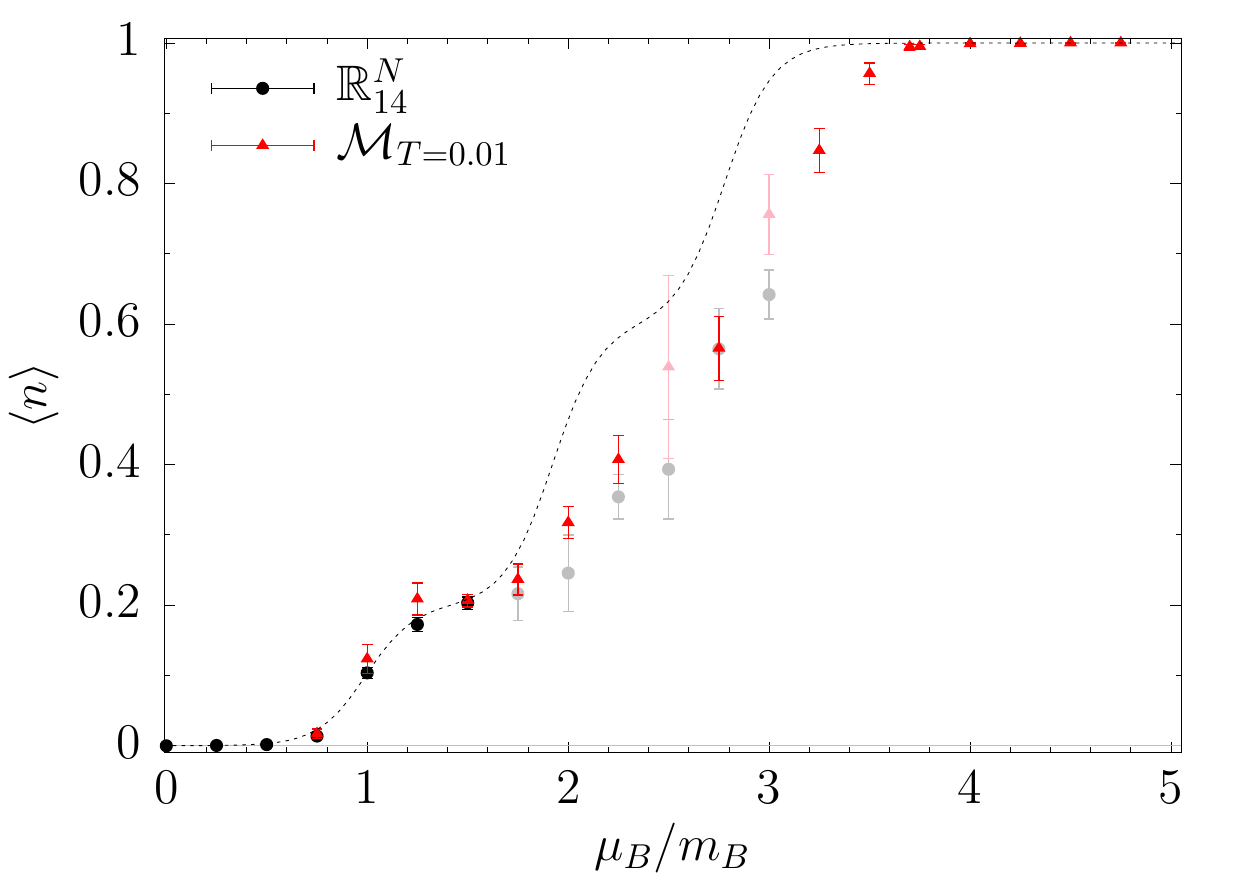}
 \includegraphics[width=0.48\linewidth]{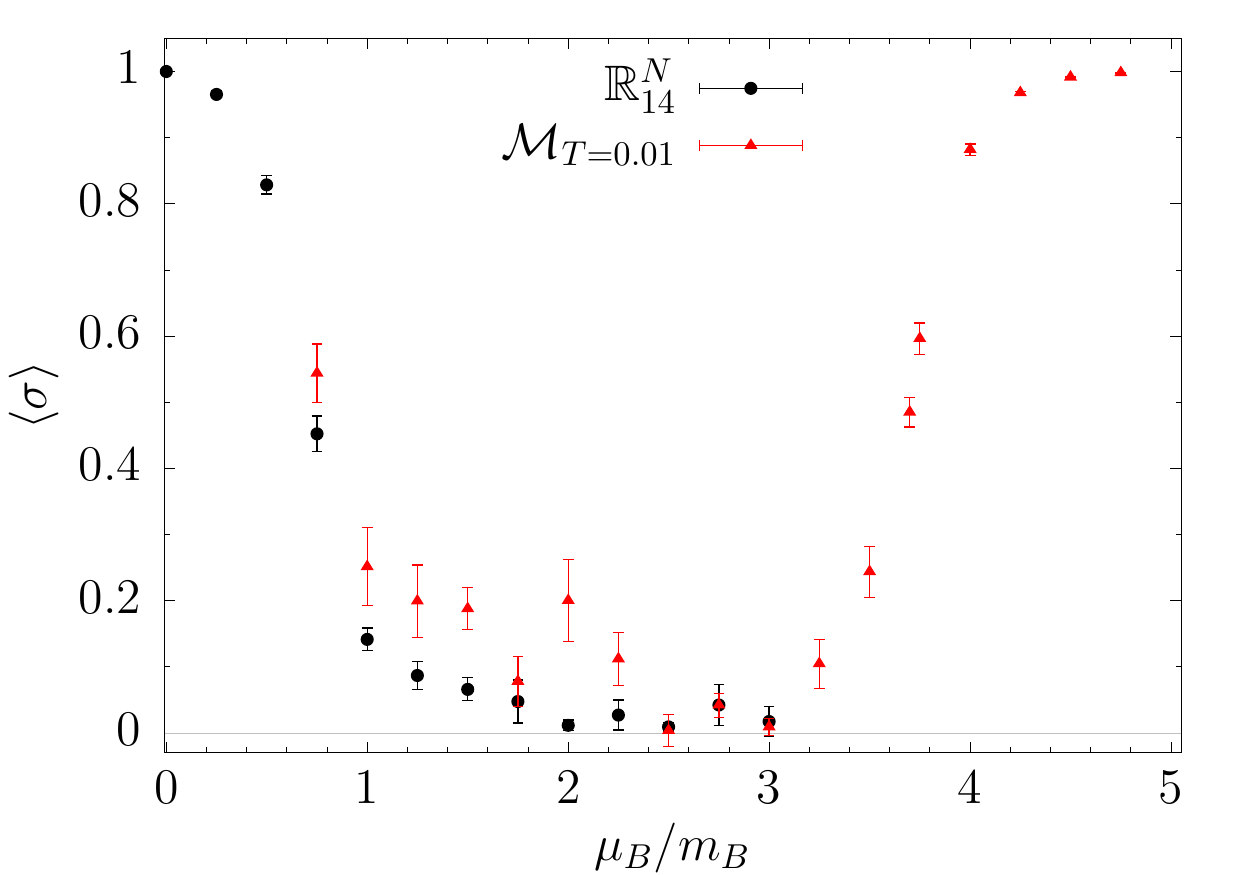}
\caption{\label{fig:1410}Density $\langle n\rangle$ and average sign $\av{\sigma}$ as a function of $\mu_B/m_B$ for staggered fermions on a lattice of size $14\times10$.  The dashed curve represents the free baryon gas with the same mass.}
\end{figure*}

The bare parameters $g=0.5$ and $m=0.05$ yield the renormalized baryon mass $a m_B\approx 0.6$, below the lattice cutoff scale.  We have undertaken calculations on a fixed spatial lattice size $n_x=10$ at three different inverse temperatures $n_t=6,10,14$.  The fermion density and $\langle \sigma\rangle$ are presented for a $10\times10$ lattice in Fig.~\ref{fig:1010}.   Consistent results for the density are found, while $\as$ increases with larger $T$.  On the colder lattice, 14$\times$10, one sees the {\em Silver Blaze} phenomenon at small $\mu_B/m_B$ in Fig.~\ref{fig:1410}, as well as development of a plateau at the one baryon threshold.  Sample $h_T/h_0$ for the $n_t=10$ lattice are in Table~\ref{tab:ht}.  $h_T$ for some $\mu_B/m_B$ exceeds unity on both lattices, implying flow reduces the computational time.

\begin{table}
 \begin{center}
  \begin{tabular}
   {l| c c c c}
   \hline\hline
   $T$ & $t_{\rm config,T}/t_{\rm config,0}$ & $\Sigma_T/\Sigma_0$ & $\langle\sigma\rangle_T/\langle\sigma\rangle_0$ & $h_T/h_0$  \\\hline
   0.01 & 6&0.8&3&1.2\\
   0.02 & 18&0.7&4&0.6\\
   0.05 & 28&0.5&13&3\\
    \hline
    \hline
  \end{tabular}
    \caption{\label{tab:ht}Maximum figure of merit $h_T/h_0$ for different flow times $T$ measured on the $10\times10$ lattice.}
 \end{center}
\end{table}

GTM may be applied to gauge theories without encountering any fundamental obstacles, as shown in $QED_{1+1}$. The difficulties that plague the definition of Lefschetz thimbles in a gauge theory are found to be innocuous.  Despite the holomorphic flow requiring greater computational time per configuration compared to the real plane, the improvement in the average sign reduces the time needed to compute observables at a fixed precision compared to an equivalent real plane calculation. 
 
\begin{acknowledgments}
The work was completed in collaboration with Andrei Alexandru, Gokce Basar, Paulo Bedaque, and Scott Lawrence, all of whom the author would like to thank.  H.L. is supported by U.S. Department of Energy under Contract No. DE-FG02-93ER-40762.
\end{acknowledgments}

\bibliographystyle{JHEP}
\bibliography{thimbology}

\providecommand{\href}[2]{#2}\begingroup\raggedright\begin{thebibliography}{10}

\bibitem{Cristoforetti:2012su}
{\scshape AuroraScience}, M.~Cristoforetti, F.~Di~Renzo and L.~Scorzato,
  \emph{{New approach to the sign problem in quantum field theories: High
  density QCD on a Lefschetz thimble}},
  \href{https://doi.org/10.1103/PhysRevD.86.074506}{\emph{Phys. Rev.}
  {\bfseries D86} (2012) 074506}
  [\href{https://arxiv.org/abs/1205.3996}{{\ttfamily 1205.3996}}].

\bibitem{Cristoforetti:2013wha}
M.~Cristoforetti, F.~Di~Renzo, A.~Mukherjee and L.~Scorzato, \emph{{Monte Carlo
  simulations on the Lefschetz thimble: Taming the sign problem}},
  \href{https://doi.org/10.1103/PhysRevD.88.051501}{\emph{Phys. Rev.}
  {\bfseries D88} (2013) 051501}
  [\href{https://arxiv.org/abs/1303.7204}{{\ttfamily 1303.7204}}].

\bibitem{Cristoforetti:2013qaa}
M.~Cristoforetti, F.~Di~Renzo, A.~Mukherjee and L.~Scorzato, \emph{{Quantum
  field theories on the Lefschetz thimble}}, {\emph{PoS} {\bfseries
  LATTICE2013} (2014) 197} [\href{https://arxiv.org/abs/1312.1052}{{\ttfamily
  1312.1052}}].

\bibitem{Scorzato:2015qts}
L.~Scorzato, \emph{{The Lefschetz thimble and the sign problem}}, {\emph{PoS}
  {\bfseries LATTICE2015} (2016) 016}
  [\href{https://arxiv.org/abs/1512.08039}{{\ttfamily 1512.08039}}].

\bibitem{Alexandru:2015sua}
A.~Alexandru, G.~Basar, P.~F. Bedaque, G.~W. Ridgway and N.~C. Warrington,
  \emph{{Sign problem and Monte Carlo calculations beyond Lefschetz thimbles}},
  \href{https://doi.org/10.1007/JHEP05(2016)053}{\emph{JHEP} {\bfseries 05}
  (2016) 053} [\href{https://arxiv.org/abs/1512.08764}{{\ttfamily
  1512.08764}}].

\bibitem{Alexandru:2017czx}
A.~Alexandru, P.~F. Bedaque, H.~Lamm and S.~Lawrence, \emph{{Deep Learning
  Beyond Lefschetz Thimbles}},
  \href{https://doi.org/10.1103/PhysRevD.96.094505}{\emph{Phys. Rev.}
  {\bfseries D96} (2017) 094505}
  [\href{https://arxiv.org/abs/1709.01971}{{\ttfamily 1709.01971}}].

\bibitem{Mori:2017pne}
Y.~Mori, K.~Kashiwa and A.~Ohnishi, \emph{{Toward solving the sign problem with
  path optimization method}},
  \href{https://doi.org/10.1103/PhysRevD.96.111501}{\emph{Phys. Rev.}
  {\bfseries D96} (2017) 111501}
  [\href{https://arxiv.org/abs/1705.05605}{{\ttfamily 1705.05605}}].

\bibitem{Alexandru:2018fqp}
A.~Alexandru, P.~F. Bedaque, H.~Lamm and S.~Lawrence, \emph{{Finite-Density
  Monte Carlo Calculations on Sign-Optimized Manifolds}},
  \href{https://arxiv.org/abs/1804.00697}{{\ttfamily 1804.00697}}.

\bibitem{Mori:2017nwj}
Y.~Mori, K.~Kashiwa and A.~Ohnishi, \emph{{Application of a neural network to
  the sign problem via the path optimization method}},
  \href{https://doi.org/10.1093/ptep/ptx191}{\emph{PTEP} {\bfseries 2018}
  (2018) 023B04} [\href{https://arxiv.org/abs/1709.03208}{{\ttfamily
  1709.03208}}].

\bibitem{Bursa:2018ykf}
F.~Bursa and M.~Kroyter, \emph{{A simple approach towards the sign problem
  using path optimisation}},
  \href{https://arxiv.org/abs/1805.04941}{{\ttfamily 1805.04941}}.

\bibitem{Alexandru:2018ddf}
A.~Alexandru, P.~F. Bedaque, H.~Lamm, S.~Lawrence and N.~C. Warrington,
  \emph{{Fermions at Finite Density in (2+1)d with Sign-Optimized Manifolds}},
  \href{https://arxiv.org/abs/1808.09799}{{\ttfamily 1808.09799}}.

\bibitem{Schmidt:2017gvu}
C.~Schmidt and F.~Ziesche, \emph{{Simulating low dimensional QCD with Lefschetz
  thimbles}}, {\emph{PoS} {\bfseries LATTICE2016} (2017) 076}
  [\href{https://arxiv.org/abs/1701.08959}{{\ttfamily 1701.08959}}].

\bibitem{Alexandru:2018ngw}
A.~Alexandru, G.~Basar, P.~F. Bedaque, H.~Lamm and S.~Lawrence, \emph{{Finite
  Density $QED_{1+1}$ Near Lefschetz Thimbles}},
  \href{https://doi.org/10.1103/PhysRevD.98.034506}{\emph{Phys. Rev.}
  {\bfseries D98} (2018) 034506}
  [\href{https://arxiv.org/abs/1807.02027}{{\ttfamily 1807.02027}}].

\end{thebibliography}\endgroup

\end{document}